\documentclass{article}

\usepackage{PRIMEarxiv}

\usepackage[utf8]{inputenc} 
\usepackage[T1]{fontenc}    
\usepackage{hyperref}       
\usepackage{url}            
\usepackage{booktabs}       
\usepackage{amsfonts}       
\usepackage{nicefrac}       
\usepackage{microtype}      
\usepackage{lipsum}
\usepackage{fancyhdr}       
\usepackage{graphicx}       
\graphicspath{{media/}}     

\usepackage{booktabs}
\usepackage{amsmath}
\usepackage{amssymb}
\usepackage{natbib}
\usepackage{xcolor}
\usepackage{multirow}
\usepackage{bbm}
\usepackage[caption=false]{subfig}
\def\reals{\mathbb{R}}

\renewcommand\vec{\mathbf}

\title{Phenotyping with Positive Unlabelled Learning for Genome-Wide Association Studies
\thanks{\textit{\underline{Citation}}: 
\textbf{Authors. Title. Pages.... DOI:000000/11111.}} 
}

\author {
    Andre Vauvelle,\textsuperscript{\rm 1}
    Hamish Tomlinson, \textsuperscript{\rm 2}
    Aaron Sim, \textsuperscript{\rm 2}
    Spiros Denaxas, \textsuperscript{\rm 1}  \\
     \textsuperscript{\rm 1} Institute of Health Informatics, University College London, UK\\
    \textsuperscript{\rm 2} Benevolent AI, 4-8 Maple St, London, UK\\
}

\begin{document}
\maketitle

\begin{abstract}
Identifying phenotypes plays an important role in furthering our understanding of disease biology through practical applications within healthcare and the life sciences. The challenge of dealing with the complexities and noise within electronic health records (EHRs) has motivated applications of machine learning in phenotypic discovery. While recent research has focused on finding predictive subtypes for clinical decision support, here we instead focus on the noise that results in phenotypic misclassification, which can reduce a phenotypes ability to detect associations in genome-wide association studies (GWAS). We show that by combining anchor learning and transformer architectures into our proposed model, AnchorBERT, we are able to detect genomic associations only previously found in large consortium studies with 5$\times$ more cases. When reducing the number of controls available by 50\%, we find our model is able to maintain 40\% more significant genomic associations from the GWAS catalog compared to standard phenotype definitions.
\end{abstract}

\section{Introduction}
\label{sec:intro}

As the collection of healthcare data has expanded, traditional definitions of disease have been challenged due to large differences in outcomes between patients. \emph{Phenotyping} refers to the process of defining  a clinically relevant set of characteristics, such as exposures and outcomes for the purpose of patient identification. These characteristics can include simple traits, such as eye colour, but also extend to include definitions of disease from as wide as diseases of the circulatory system to specific disease subtypes. Identifying these phenotypes plays an important role in furthering our understanding of disease for applications within epidemiological research and drug development.  

\begin{figure*}[htbp]
\centering
  \includegraphics[width=\linewidth]{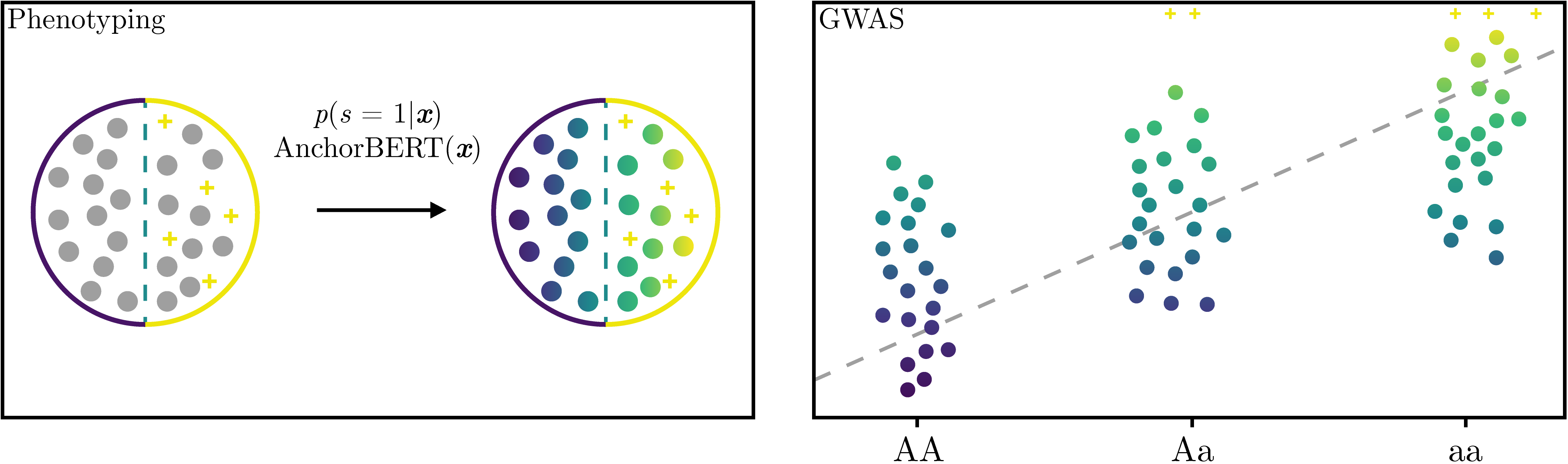}
  \label{fig:overview}
  \caption{Overview of AnchorBERT phenotyping for GWAS. Unlabelled patients (grey dots) are given predicted probabilities of having the anchor variable (yellow plusses) by AnchorBERT. These patients are then used as a continuous trait in linear regression GWAS.}
\end{figure*}



In the first approach, panels of experts define phenotypes by a series of rules \citep{denaxas_uk_2019}. While they are based on a consensus of domain experts, the scalability of rule-based methods are limited in that they are laborious, iterative, and time consuming processes \citep{banda_advances_2018}. As electronic health records grow, there is an opportunity to conduct large-scale analyses to drive our understanding of disease biology, however this will make expert review infeasible. Thus, we turn to machine learning.

Machine learning has been previously used to identify phenotypes from electronic health records, primarily in the context of clinical decision support. For instance, in \citet{miotto_deep_2016, zhang_data-driven_2019}, phenotypes are pragmatically identified to enable particular tasks such as characterising patients with a likelihood to require more specialised care, or are at risk of deterioration and/or death. Evaluation of machine learning generated phenotypes has focused on predictive measures of patient outcomes and response to treatment. This makes sense in a clinical setting where phenotypes need to be predictive of the future health state of a patient \citep{lee_temporal_2020}. 


In this paper, we focus on the phenotyping task of \emph{diagnosis classification} in electronic health records for \emph{biological discovery}. Traditionally, phenotype labels are assigned to patients according to the presence or absence of International Classification of Diseases (ICD) codes. However, this method is likely to mislabel due to the large amount of noise and complexity within EHR data. Previous studies have shown that identifying phenotypes using diagnostic codes as a proxy can result in poor positive and negative predictive values across diseases and healthcare systems \citep{woodfield_accuracy_2015}. Machine learning approaches may enable us to make fewer assumptions about the fidelity of individual codes while learning from longitudinal patient histories. Concretely, our goal is to learn to classify patients according to whether or not they have a disease.



\begin{table}[bp]
\caption{Diseases studied}
\centering
\begin{tabular}{lll}
\toprule
Acronym & Disease & Phecode \\
\midrule
MI & Myocardial Infarction & 411.1  \\
T2D & Type 2 Diabetes & 250.2 \\
HF & Heart Failure & 428.2 \\
DM & Dementia & 290.1 \\
RA & Rheumatoid Arthritis & 714.0$\vert$714.1\\
\bottomrule
\end{tabular}
\label{tab:diseases}%
\end{table}

Within diagnosis classification, we identify two distinct issues:
1) Heterogeneity. Current definitions of disease are too broad such that multiple distinct phenotypes exist that better describe the presenting patient. In the context of biological discovery, conflicting influences from multiple subtypes can reduce p-values and distort effect sizes in association studies. 2) Phenotypic misclassification. This instead considers the possibility that diseases have been either incorrectly identified or missed during clinical observation and data collection. This noise can result in mislabelling of cases and controls, reducing power in association studies. Incorrectly assigning a patient as a case occurs when a patient is misdiagnosed as having the desired phenotype. Conversely, identifying controls with an absence of a diagnosis does not necessarily mean the diagnosis was ruled out. Instead, the diagnosis may have been missed. While the first issue has previously been addressed with subtyping \citep{li_identification_2015}. We instead focus on the second issue of phenotypic misclassification, which has received comparatively less attention.

We propose a robust metric for analysing phenotyping algorithms in the context of biological discovery. Our approach is to report replicated associations from previous studies in the GWAS catalog \citep{buniello_nhgri-ebi_2019}. Importantly, Genome-wide association studies present a unique means of identifying phenotypes with distinct causal disease biology \citep{dahl_genetic_2020}. GWAS often require high sample sizes, particularly when effect sizes may be small, and indeed many GWAS catalog associations have been contributed by large consortia. Alternatively, we attempt to find associations that meet significance thresholds by reducing the noise due to misclassification and thus improve effective effect size. Evaluating phenotypes in this manner presents a more relevant metric to biological discovery, an independent and robust alternative to outcome-based evaluation.

\subsection*{Problem Specification}

\textbf{The main aim of this work} is to create robust phenotypes for genomic discovery by addressing the issue of phenotypic misclassification. 

Our methodological contributions combine two areas of research; transformer architectures and anchor variable models. We reduce phenotypic misclassification within EHRs by treating the data as only positive and unlabelled data. Specifically, we use an anchor variable model to predict the probability of a patient being a case, employing a transformer model to improve the approximation of this probability in a model we call AnchorBERT. Finally, we validate our models against current diagnosis classification methods by reproducing known associations from five different diseases using a repository of validated studies in the GWAS catalog. Overall, our contributions are summarised as:

\begin{itemize}
    \item We introduce the distinct issue of phenotypic misclassification and present the first model using noisy label learning and state-of-the-art deep learning architectures to improve genomic associations for biological discovery.    
    
    \item We present a robust, independent validation metric, more suited to biological discovery, based on replicating genomic associations found in the GWAS catalog.
    
    \item Our proposed AnchorBERT model outperforms standard phenotypes definitions by maintaining more known associations as the number of samples used in GWAS is reduced. We are able to reproduce genetic associations only previously found in large consortium studies.
    
\end{itemize}

\section{Related Work}

Previous works have looked to improve GWAS power for diseases with poor positive predictive value by setting an increased threshold on the number of total diagnosis codes required to be classified as a case \citep{diogo_phenome-wide_2018}. \citet{sinnott_pheprob_2018} show that setting a threshold to define a case can be avoided entirely by instead modelling the probability of a phenotype with unsupervised clustering (Pheprob). This continuous probability is then used directly in the association study in place of a binary classification, which \citet{sinnott_pheprob_2018} shows improved power to detect associations. Pheprob includes the total number of codes as an additional variable as part of a parametric binomial mixture model. The method provides some ability to assign non-zero probabilities to controls but largely ignores the possibility of false-negative labels.

Mislabelling controls has a smaller effect on power \citep{edwards_power_2005}, yet determining which error rate is larger is difficult and largely unknown \citep{zwaan_diagnostic_2020}. Previous studies have considered treating cases as positive only and control data as unlabelled data \citep{halpern_using_2014, agarwal_learning_2016}. Rather than updating the probability of controls for use in association studies, these studies focus on semi-automated methods for improving phenotype definitions by iteratively updating anchor variables. 

Previous structured EHR machine learning works \citep{si_deep_2020} suggest that models able to capture non-linear and sequential features between past, present, and future events produce better results on predictive tasks. Although our task is ultimately aimed at finding genetic associations, we hypothesise that more predictive representations will allow greater identification of noisy samples, reducing their negative influence on the control cohort and, in turn, provide greater overall associative power. \citet{hansen_towards_2018} applied autoencoders with anchor learning but did not evaluate performance to detect associations. \citet{yu_enabling_2018} compare unsupervised clustering using mixture model and anchor learning methods but do not reproduce genomic associations and heavily rely on clinical notes.

\section{Methods}

\subsection{Problem Formulation}

Let $\vec{x}=\{x_t\}^T_{t=1}$ describe the sequential collection of disease codes from a total of $T$ visits in a patient's health record. Each visit $x_t \in \mathcal{X}$, is a multi-hot encoding of the $d$-total diseases, such that $x_{t, j}$ is marked as $1$ if the $j^\text{th}$ disease was observed during visit $t$ else $0$. We let $y \in \reals^d$ describe the latent disease state; this describes all diseases occurring during a patients life including those that are not observed in $\vec{x}$. Finally, let $g$ denote the genetic variable of interest measured for a patient, such as a SNP taking a value in $\{0,1,2\}$. For each patient, we consider the collection $\{\vec{x}, y; g\}$.

Ideally, we would be able to measure the association between the true, latent, disease state $y$ and genetic variables $g$ by defining a case-control cohort on the presence of the disease state; however, we are only presented with the observed diseases $\vec{x}$. For GWAS, research typically proceeds by first labelling disease cases ($y_j = 1$) if any $x_{t, j} = 1$ for $t \in \{1, \ldots, T\}$. Similarly controls ($y_j = 0$) if all $x_{t, j} = 0$ for $t \in \{1, \ldots, T\}$. From now on, we assume that we are only interested in the latent disease $j=a$ and drop the index from $y_j$ such that $y\in \{0,1\}$.

We propose learning the function $p(y=1|\vec{x}) = f(\vec{x})$ and using this as a continuous trait in a regression to detect associations with $g$.

\subsection{Anchor learning}
Anchor learning is a previously well-studied method that can be applied to only positive and unlabelled data \citep{elkan_learning_2008, halpern_using_2014}. Here, we reproduce the fundamental formalisation with reference to our task.

We frame our problem as having positive only and unlabelled data by assuming that observing a specific disease code only positively identifies the latent class it is supposed to measure. For any patient without this code, we cannot say if they have the latent disease $y$ or not.

More generally, let $s$ indicate if the patient's sequence $\vec{x}$ is labelled, such that we know the value of $y$. $\vec{x}$ is positively labelled with $y=1$ if $s=1$. If $s=0$ then the label of $\vec{x}$ is unknown, it could take either of the values, $y=0$ or $y=1$. We are assuming that only positive examples are labelled, which can be stated as 
\begin{equation}
    p(s = 1| \vec{x}, y = 0) = 0.
    \label{eq:pos_label}
\end{equation}

Our goal is to model $p(y = 1 | \vec{x})$. However, as stated above, we are assuming we only have positive labelled patients. \citet{elkan_learning_2008} show it is possible to progress if we assume our positive examples are chosen randomly from the set of all positive examples. This can also be stated by saying that $s$ and $\vec{x}$ are conditionally independent given $y$ or 
\begin{equation}
    p(s=1\vert \vec{x},y=1) = p(s=1\vert y=1). \label{eq:conditional_independent}
\end{equation}

Using this assumption, we can move closer to our goal of approximating $p(y=1\vert \vec{x}$) by learning a classifier to predict our anchor variable $p(s=1|\vec{x})$ as
\begin{align}
    p(s = 1 \wedge y=1\vert \vec{x}) &= p(s=1|\vec{x}) \\
    p(y=1|\vec{x}) &= \frac{p(s=1 \vert \vec{x})}{p(s=1\vert y=1,\vec{x})}.
    \label{eq:goal}
\end{align}

More specifically, in our case $s$ is determined by the presence of a diagnosis code, $x_{t,a} = 1$ for some $t$. If disease code $a$ is present, we also we strong believe the patient has its described latent disease, $y=1$. We say $a$ is an \emph{anchor variable} where
\begin{align}
    p(s = 1 \wedge y=1\vert \vec{x}) = 1, \; & \text{if } a \in \vec{x}.
    \label{eq:anchor}
\end{align}

Considering Equation~\ref{eq:anchor} and Equation~\ref{eq:conditional_independent}, our goal can be updated to
\begin{equation}
    p(y=1|\vec{x}) = 
    \begin{cases}
     1,            & \text{if } a \in \vec{x} \\
     \frac{p(s=1 \vert \vec{x})}{c}, & \text{if } a \notin \vec{x}.
     \label{eq:anchor_goal}
\end{cases}
\end{equation}
where $c = p(s=1 \vert y=1)$ is a constant. This means we can learn an \emph{anchor classifier}, $h(\vec{x}) = p(s=1 \vert \vec{x})$, in place of $f(\vec{x})$ to rank our instances when $a \notin \vec{x}$. When the anchor variable is present in the patient in the patients’ data, the probability of the latent disease is 1. Otherwise, the probability is given by the result of an anchor classifier.

Comparing Equation~\ref{eq:anchor_goal} to the traditional definition of cases and controls, it is possible to view our task as learning from noisy labels. We are assuming that the standard definition of a control is noisy and, instead, learn a model to assign controls with scores that are higher if $\vec{x}$ is predictive of being a case.

\begin{table*}[htbp]
  \caption{Mean and standard deviation anchor variable classifier model performance on test set across each investigated disease. AUPRC: Area under precision recall curve, AUROC: Area under receiver operator curve, LR: Logistic Regression.}
\centering
    \begin{tabular}{llllll}
    \toprule
         &  & \multicolumn{2}{l}{AUPRC} & \multicolumn{2}{l}{AUROC} \\
    Disease & \# Cases (ratio) & LR & BERT &                 LR & BERT \\
     \cmidrule(r){1-2}  \cmidrule(r){3-4}  \cmidrule(r){5-6}
    MI & 18,007 (.0577)  & .5547 $\pm$ .0000 & \textbf{.6680 $\pm$ .0021} & .9039 $\pm$ .0000 & \textbf{.9538 $\pm$ .0006}\\
    T2D & 31,801 (.1019) & .4374 $\pm$ .0000 & \textbf{.5071 $\pm$ .0012} & .7980 $\pm$ .0001 & \textbf{.8386 $\pm$ .0009}\\
    HF & 9,179 (.0294)   & .3806 $\pm$ .0000 & \textbf{.4453 $\pm$ .0043} & .9208 $\pm$ .0000 & \textbf{.9405 $\pm$ .0013}\\
    DM & 4,582 (.0147)   & .2380 $\pm$ .0000 & \textbf{.3286 $\pm$ .0101} & .8507 $\pm$ .0000 & \textbf{.8842 $\pm$ .0024}\\
    RA & 7,956 (.0255)   & .1029 $\pm$ .0000 & \textbf{.1375 $\pm$ .0031} & .7603 $\pm$ .0000 & \textbf{.8037 $\pm$ .0016}\\
    \bottomrule
    \end{tabular}
  \label{tab:anchor_metrics}%
\end{table*}

\subsection{Using BERT as anchor classifier}

Previously \citet{halpern_electronic_2016} used a logistic regression model, similar to $h(\vec{x};\theta) = \sigma\left(\sum_j (\sum_t x_{t,j}) \theta_j\right)$ within anchor learning to update the terms of phenotype definitions. 

In addition to instead using the output of the anchor classifier directly in downstream GWAS, we propose AnchorBERT, which combines NLP-like embeddings and the encoder of a transformer model to model $h(\vec{x})$. This model is inspired by the original BERT model within the NLP domain \citep{devlin_bert_2019}. 

\subsubsection{Multi-head attention}

The main component of our model relies upon entirely on (self)-attention mechanisms \citep{vaswani_attention_2017}. The primary advantage being that we are able to model global dependencies between codes, $x_t$, regardless of their distance from each other in the sequence $\vec{x}$.



Self-attention modules first associate each input value $x_t$ with a query $q$ and a set of key-value pairs ($k$, $v$), where the queries, keys and values are themselves linear projections of the input:

\begin{align}
    q=\theta^{q}x_t, &&
    k=\theta^{k}x_t, &&
    v=\theta^{v}x_t  &&
\end{align}
where $\theta^{q}$, $\theta^{k}$ and $\theta^{v}$ are to be learnt. A self-attention score then determines how much focus to place on each output value given $x_t$ compute with the dot-product of the queries and keys:

\begin{equation}
    \text{Attention}(Q, K, V) = \text{softmax}\left(\frac{Q K^{T}}{\sqrt{k}}\right)V
    \label{eq:attention}
\end{equation}
where each line of Q (resp. K) are matrices whose rows are the values associated with each query (resp. keys) entries. V is the matrix whose rows are the values associated with each input data. Our model uses multi-head self-attention where the input is independently processed by $n$ self-attention modules. This leads to $n$ outputs which are concatenated back together to form the final attention output vector. For further details and clarification, see \citet{vaswani_attention_2017}.

\subsubsection{Embeddings}

Since self-attention does not use any recurrent or convolutional mechanisms, we need to include an additional encoding to provide information on the absolute and relative position of inputs. We follow \citet{li_behrt_2020} by including a unique predetermined positional embedding for each visit. Segmentation embeddings are also included; these are two trainable vectors that alternate between subsequent visits and provide additional flexibility to encode differences between visits. 

Each term $x_{t,j}$ uses a learnt embedding to convert from a one-hot vector of dimension $d$ to $d_{model}$. Since our model can attend to each term within a visit equally, this is equivalent to the unordered multi-hot representation in our formulation.

An attention mask, $A$, is used on individual anchor disease terms in the patient sequence. This is effectively equivalent to removing them entirely and negates any influence during training and evaluation. Equation~\ref{eq:attention} becomes  
\begin{equation}
    \text{Attention}(Q, K, V) = \text{softmax}\left(\frac{Q K^{T}}{\sqrt{k}} + A \right)V,
\end{equation}
where the elements of the attention mask, $A$, are zero except for at positions corresponding to $x_t=a$, where a large negative value is used.

Tokenization of the disease codes occurs before being fed as input to our model. Unique tokens are assigned to each term with a total count greater than 0.01\% of the total terms; otherwise, a [UNK] token is used. [SEP] tokens are added between each patient stay (episode). [PAD] tokens are appended to each sequence to maintain a maximum length of 256. [CLS] tokens are added at the start of each patient record for the BERT prediction pooling scheme.


\subsubsection{Training and Prediction}

Our model is updated during training by minimizing binary cross-entropy loss. The binary labels are positive if there are any anchor variables present in a patients data. We make anchor variable predictions for an entire sequence by simply taking the final attention output vector corresponding to the [CLS] token and applying a linear layer. The sigmoid function is applied to the logits from the linear layer to produce the predicted anchor probabilities. 

Once training and evaluation are complete, we can produce the final phenotype probabilities $p(y=1|\vec{x})$. As shown in Equation~\ref{eq:anchor_goal}, we replace the predicted anchor probability with $1$ if the anchor label is positive. We set $c=1$ as we only need to rank our examples according to the chance that they belong to class $y = 1$ \citep{elkan_learning_2008}.


\subsection{Baselines}

We first compare AnchorBERT to the previously studied anchor logistic regression (Anchor LR). We apply logistic regression by aggregating our sequence of disease tokens into total counts and scaling to a unit normal distribution. Scaled anchor counts are removed from the input features. We use the sklearn implementation of logistic regression with the L-BFGS-B solver. 

We also compare our anchor variable models against the commonly used thresholding method and Pheprob baseline \citep{sinnott_pheprob_2018}. The thresholding models produce a binary phenotype, classifying a patient as a case if the total number of anchor phecodes equals or exceeds a threshold. We use the original implementation of Pheprob, a parametric binomial mixture model that includes the total count of all a patient's phecodes and anchor phecodes as input features. Pheprob outputs a continuous phenotype probability for each patient.

\subsection{Anchor Performance Metrics}
Anchor models are evaluated using the area under the receiver-operator curve (AUROC) and precision-recall curve (AUPRC). Average precision is used for AUPRC. While both metrics assess a model's ability to predict the anchor variable, AUPRC removes the influence of true-negative controls allowing us better to assess the positive predictive power of the models.

\subsection{Hyperparameters}

Hyperparameters for all experiments are detailed the Appendix. Anchor classifiers are trained using a training:validation:test split of 6:2:2. Test data is unseen until final evaluation, and hyperparameters are tuned via grid search optimising for validation AUPRC. Hyperparameters are fitted for each disease, then refitted with a new seed for each evaluation. The best performing hyperparameters for each disease are shown in Table~\ref{tab:bert_hparams}.

Dropout is applied to the multi-head attention layers and within the final linear classification head. The model binary cross-entropy loss is optimised with BertAdam \cite{devlin_bert_2019}. We set the size of embeddings and linear classification heads to be determined by a shared hidden size hyperparameter. See our full PyTorch Lightning implementation based on the Hugging Face \citep{xia_part-dependent_2020} BERT model at \url{https://github.com/andre-vauvelle/AnchorBERT} for clarification.

\section{Experiments and Results}
In this section, we provide the details and results of our experiments on the UK Biobank data. We first introduce the UK Biobank data, the diseases studied, and any data preprocessing requirements. Second, we report the performance of our anchor variable models and describe an experiment to investigate their robustness to control noise and compatibility with our PU data assumption. Finally, we can compare AnchorBERT to our baseline phenotype models on their ability to reproduce known genetic associations.

\begin{table*}[b]
\caption{Basic statistics of UK Biobank EHR data after preprocessing.}
\centering
\begin{tabular}{|ll||ll|}
\# of unique   patients & 321,837 & \# of Phecodes & 16,655,024 \\
\# of visits &  7,698,687  & Avg. \# of Phecodes per visit & 2.16 \\
Avg. \# of visits per patient & 23.9 & Maximum sequence length & 256 \\
Avg. \# of Phecodes per patient & 51.7 & \# of unique Phecode tokens & 837
\end{tabular}
\label{tab:basic_stats}
\end{table*}

\subsection{UK Biobank data}
The UK Biobank \citep{sudlow_uk_2015} is a national population-based study comprising of 502,629 individuals. We extract all available diagnosis terms from both primary (Read V2 and V3) and secondary (ICD-10) care settings for every patient. We then map raw terms to Phecodes using the previously validated Phecode map 1.2b from \citet{wu_mapping_2019}. We use Phecodes as our disease terms since the extracted raw terms are often distinguished by billing-specific information. Here Phecodes provide a higher level, clinically meaningful traits more suitable for genetic association studies \citep{denny_phewas_2010}. All unmappable terms are dropped. Finally, only patients with $\geq5$ terms are retained, resulting in a total of 321,837 patients. We study five different diseases, identified by their Phecode and detailed in Table~\ref{tab:diseases}.

Genomic data extraction and quality control processing follows the same methodology as \citet{garfield_relationship_2021}. This data is linked to all possible phenotyped patients resulting in 312,010 patients with both genetic and phenotype data. More detailed statistics of the data are available in Table~\ref{tab:basic_stats}.

\subsection{Anchor Classifier Performance and Robustness to Control Noise}

\begin{figure*}[tbp]
\centering
\subfloat[]{
  \includegraphics[width=0.40\linewidth]{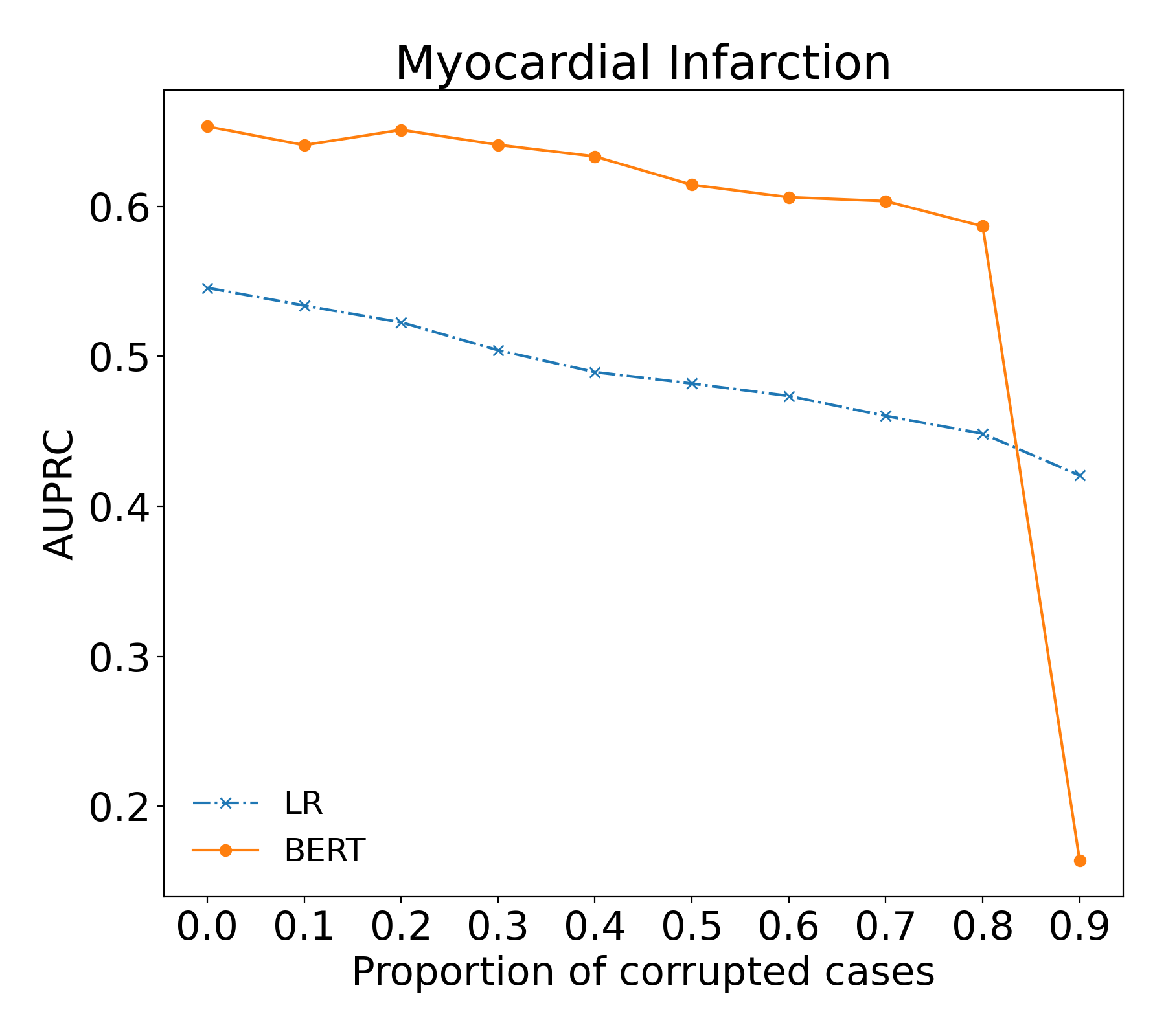}%
  \label{fig:mi_noise}
  }
  \subfloat[]{
  \includegraphics[width=0.40\linewidth]{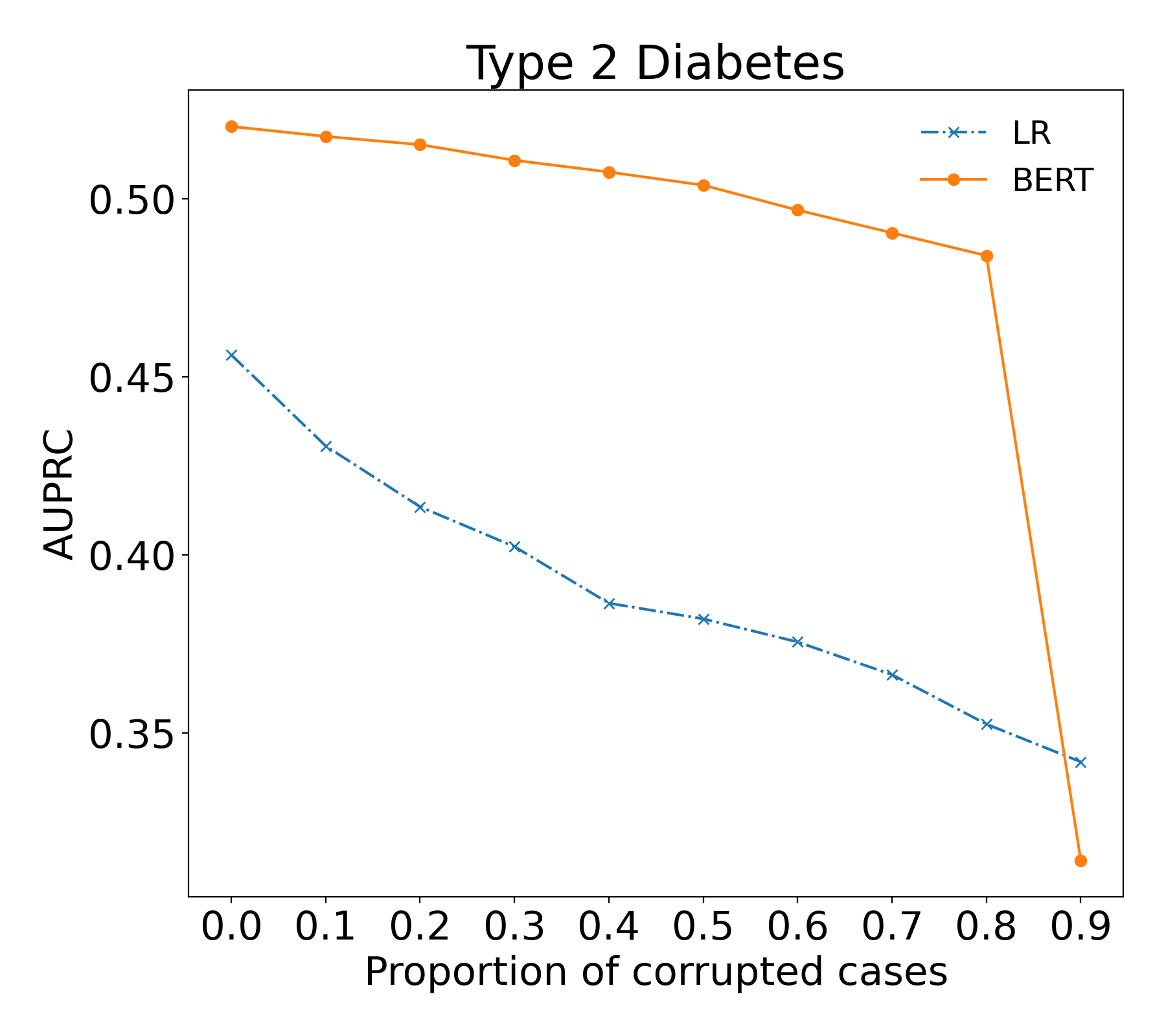}%
  \label{fig:t2d_noise}
  }
\caption{Area under precision recall curve for anchor variable classifiers as an increasing proportion of cases are added to the unlabelled set during training.}
\label{fig:noise_robust}
\end{figure*}

After finding optimal hyperparameters, we retrained and evaluated each anchor classifier across ten runs to report the mean and standard deviation of each performance metric.

In order to investigate if our anchor classifiers are robust to noisy controls, we would like to test if performance remains high while operating under corrupted anchor labels. Although we cannot identify which patients in our dataset have incorrectly been assigned without an expensive chart review, instead, we can artificially add noise. 

From the perspective of anchor learning, we consider the negative set as unlabelled. If we randomly switch a sub-sample of the positive anchor labels to unlabelled during training then evaluate on a validation set free from noise, we can effectively compare the performance of our anchor classifiers, $h(\vec{x})$. As noise increases, models that are sensitive to label noise should perform worse.

\subsubsection*{Results of Anchor Classifiers}

As seen in Table~\ref{tab:anchor_metrics}, the BERT classifier outperforms logistic regression across all disease areas. We observe considerable variation in performance between diseases, possibly indicating some diseases may have more comorbidity interactions which could help identify noisy labels. For example, T2D has almost double the number of cases compared to MI, yet performance is worse across both metrics and models.

In Figure~\ref{fig:noise_robust}, we show the results of our investigation into control noise robustness for MI and T2D, additional figures for the remaining diseases are in Appendix~\ref{apd}. Overall, BERT outperforms LR across all diseases and noise proportions, with the minor exception of DM at the highest noise level.

\subsection{Evaluating phenotypes with GWAS}

We use two experimental setups aimed at validating an improved ability to detect genetic associations. 1) Full data: We run GWAS using phenotypes generated for all available patients and compare the models' ability to reproduce known associations from the GWAS catalog. 2) Data ablation: We reduce the number of cases available in the GWAS and report which phenotyping methods are able to retain statistical significance for known associations. 

Associations are tested for using plink v2.0's generalised linear model, regressing SNPs against phenotype status \citep{chang_second-generation_2015}. Linear regression is used for the anchor variable and binomial mixture model continuous phenotypes, while logistic regression is used for binary threshold phenotypes. All regressions use the following covariates: sex, age, and 1-10 population structure principle components.

All reported significant SNPs, from both the GWAS catalog and our analysis, are filtered such that only those with p-value lower than $5\times 10^{-8}$ remain.

\begin{table*}[tbp]
\caption{Performance of each phenotyping method when reproducing known GWAS Catalog associations. MI - Myocardial Infarction, T2D - Type 2 Diabetes, HF - Heart Failure, DM - Dementia, RA - Rheumatoid Arthritis}
\centering
\resizebox{\linewidth}{!}{
\begin{tabular}{llllll}
\toprule
Phenotype Model & \multicolumn{5}{c}{Total Reproduced Catalog Genomic Associations (Proportion)} \\
& MI & T2D & HF & DM & RA \\
\midrule
AnchorBERT & \textbf{44 (0.3438)} & 266 (0.1513) & \textbf{3 (0.0714)} & 1 (1.0) & 32 (0.0814) \\
Anchor LR & 39 (0.3047) & 254 (0.1445) & 2 (0.0476)  & 1 (1.0) & 32 (0.0814) \\
\cmidrule(r){2-6}
Pheprob & 28 (0.2188) & 247 (0.1405) & 0 (0.0) & 1 (1.0) & 32 (0.0814) \\
Threshold-1 & 29 (0.2266) & \textbf{280 (0.1593)} & 0 (0.0) & 1 (1.0) & 32 (0.0814) \\
Threshold-2 & 25 (0.1953) & 237 (0.1348) & 0 (0.0) & 1 (1.0) & 32 (0.0814) \\
Threshold-3 & 18 (0.1406) & 200 (0.1138) & 0 (0.0) & 1 (1.0) & 32 (0.0840) \\
\midrule
Total Catalog Significant rsIDs & 128 & 1,758 & 42 & 1 & 393 \\
\bottomrule
\end{tabular}
}
\label{tab:catalog_gwas}%
\end{table*}

\subsubsection*{Comparison to GWAS Catalog}
In order to assess whether our overall phenotyping methods are able to increase the power of GWAS studies, we compare their ability to reproduce known significant associations from the GWAS catalog. 

It is difficult to directly report on study power as a set of true phenotype-gene associations is not possible to obtain. We follow and expand upon prior work \citep{sinnott_pheprob_2018} by replicating previously found associations. Rather than replicating a smaller number of hand-selected associations, we instead compare against all associations found for a disease in the GWAS catalog \citep{buniello_nhgri-ebi_2019}. The GWAS catalog is a widely used and freely available database of SNP-trait associations, including those from consortium studies with cohort sizes orders of magnitude larger than the UK Biobank. 

The catalog contains studies from vastly different populations and experimental procedures. Populations in the catalog may have been measured with a different sequencing array meaning that some SNPs may not be present in our data. In addition, reported loci in the GWAS catalog are often the result of fine-mapping, which keeps only the most likely causal SNP, discarding those highly correlated nearby in linkage disequilibrium (LD) \citep{slatkin_linkage_2008}. 

In order to partly address these issues, we expand significant SNPs from the GWAS catalog and the UK Biobank genomic data to include SNPs within LD. We do this using an LD reference panel with a threshold of $R^2>0.5$ from the 1000 genomes project \citep{durbin_map_2010}.

\subsubsection*{Data Ablation Study}
We also conduct a data ablation to study the influence of the anchor variable on the control cohort. Here, we are not reducing the amount of data available to train the anchor variable model. Instead, we randomly remove a proportion of the patients after training but before finding associations. If we reduce the case population defined by Threshold-1, study power should fall. For the anchor variable models, however, the influence of updating the probability of the controls should remain. At the extreme, with zero cases, only associations due to noisy samples should remain. 

Since the variation in p-values can change significantly as some instances are included or excluded between ablation thresholds, we repeat the association studies ten times and report confidence intervals. In order to reduce computational power requirements, we only test the phenotypes associations with significant SNPs found from all methods with full data and report the proportion of reproduced significant SNPs.

\subsubsection*{GWAS Results}
Table~\ref{tab:catalog_gwas} contains the results of our comparison against the known associations from the GWAS catalog. Our proposed anchor variable approach to creating continuous phenotype traits for GWAS reproduces more associations than other models for both MI and HF. Threshold-1 outperforms others on T2D, while all models find an equal number of significant associations in the catalog for DM and RA. Where the anchor variable models do not find a greater number of associations, the proportion of reproduced associations is still the same or slightly lower than Threshold-1, the current standard method used for GWAS studies.

\begin{figure*}[tb]
\centering
    \subfloat[Both cases and control\label{fig:mi_ablation_both}]{%
      \includegraphics[width=0.40\linewidth]{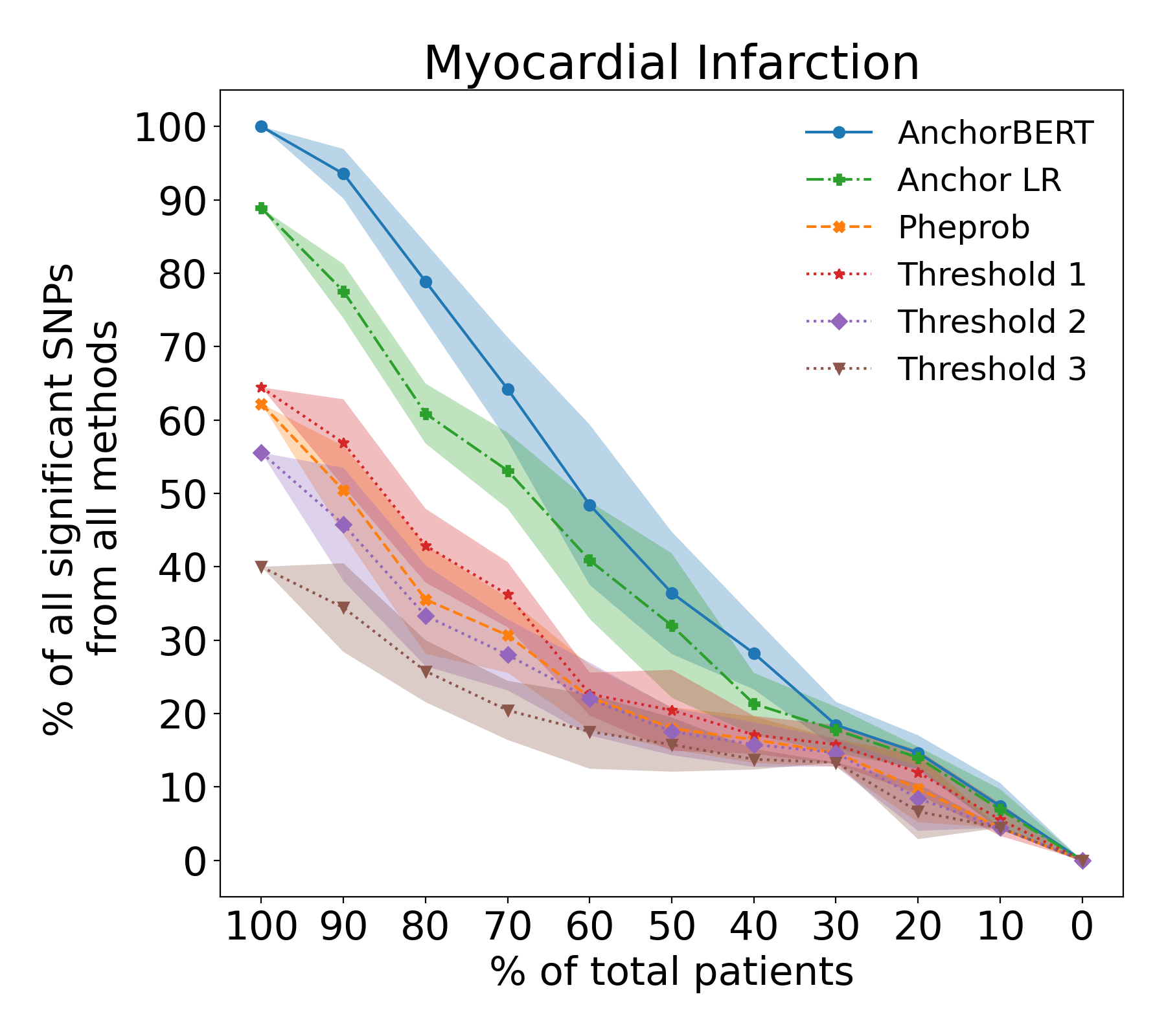}}%
    \qquad
    \subfloat[Cases only\label{fig:mi_ablation_cases}]{%
      \includegraphics[width=0.40\linewidth]{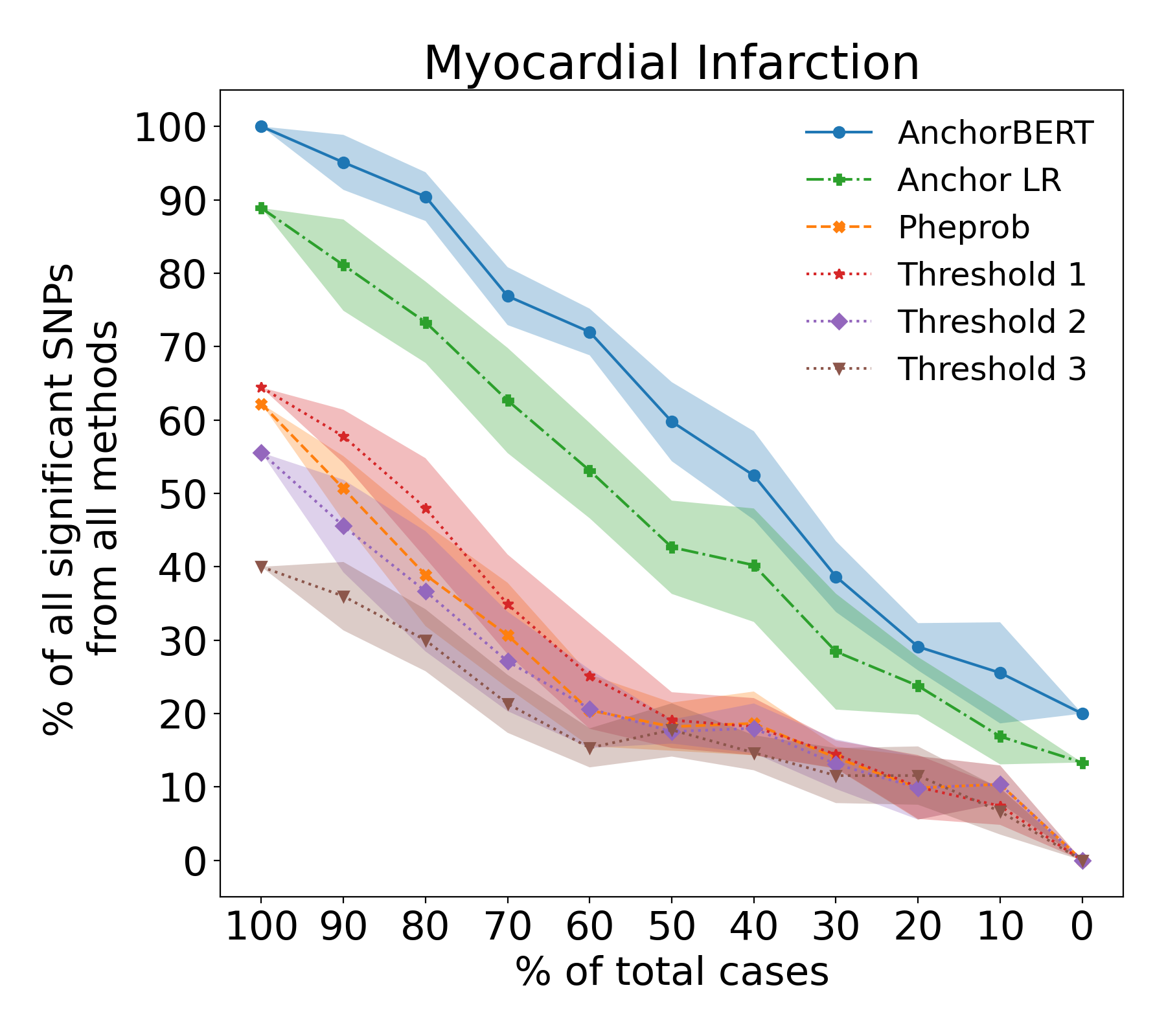}}
  
  \caption{Ablation study of patients for Myocardial Infarction and each phenotyping method. Shaded areas indicate one standard deviation of results across 10 trails.}
  \label{fig:mi_ablation}
\end{figure*}

Figure~\ref{fig:mi_ablation} shows the results of the ablation study for MI. Both proposed anchor variable models are able to reproduce the associations of the thresholding methods and Pheprob at all ablation thresholds, with AnchorBERT also outperforming the logistic regression anchor model. When removing all cases as defined by Threshold-1, non-anchor variable methods can no longer detect any associations, while AnchorBERT retains 20\% and Anchor LR retains 13\% of associations. 

In Appendix~\ref{apd} we show more complete results for each disease. Figure~\ref{fig:additional_ablation} shows the ablation study results for the other four diseases. With the exception of RA, when all Threshold-1 defined cases are removed, anchor models are still able to replicate catalog associations.


\section{Discussion}

In this work, we present a novel phenotyping method, AnchorBERT, which uses anchor learning and transformers to generate continuous phenotypes that allow for the detection of significant genomic associations with smaller cohorts. In seeking a more representative phenotype, we argue that EHR diagnoses can be treated as positive only and unlabelled data. PU data allows the application of anchor learning, where we introduce BERT as a novel modification to the anchor classifier. BERT allows greater performance when modelling the anchor variable and is more robust to label noise, modelled here by introducing additional cases into the unlabelled set.

Using data from the UK Biobank and GWAS catalog, we validate our proposed phenotyping methods, Anchor LR and AnchorBERT, together with baselines used by GWAS practitioners to detect genomic associations in HF previously only found in studies with $5\times$ the number of available cases. Using anchor phenotypes could enable the discovery of genomic associations otherwise inaccessible to studies with small cohorts. For example from Table~\ref{tab:catalog_gwas}, our proposed AnchorBERT replicates three significant SNPs (rs17042102, rs55730499, rs1556516). Which were previously only found in the largest HF GWAS meta-study to date, with approximately $5\times$ and $3\times$ as many cases and controls of European ancestry \citep{shah_genome-wide_2020}.


From Figure~\ref{fig:mi_ablation_cases} and Figure~\ref{fig:additional_ablation}, we are able detect known associations for patients without an anchor variable (0\% cases), suggesting that we are potentially identifying missed diagnoses. It is notable that, AnchorBERT reproduces at least as many or more genomic associations than Anchor LR, showing that the sequential and non-linear relationships between codes that result in higher anchor classification performance translate into improved ability to reproduce genomic associations. This trend is consistent across all disease areas considered, even for T2D where Anchor learning generally performs worse than Threshold-1. 

We note that the additional HF SNPs identified also have significant associations with related upstream comorbidities and traits, including: coronary artery disease, atrial fibrillation, and low-density lipoprotein cholesterol. This could be due to a genuine shared disease aetiology underlying these risk factors. Alternatively, these associations could be confounded and independently related to HF comorbidities.

Performance across all diseases is not guaranteed. We find anchor phenotypes are able to identify genetic associations with 0\% cases in all but RA. Considering Threshold-2 and 3 phenotypes outperform other phenotyping methods on RA, we suggest that noise in the case definition could be responsible for poor performance, as this violates our assumption in Equation~\ref{eq:anchor}.

We hope that these findings will help further efforts to discover new disease-genomic associations. Ultimately leading to a greater understanding of disease and a better, more efficient process for discovering new medicines.

\section*{Acknowledgments}
Andre Vauvelle is supported by a Benevolent AI PhD studentship. We thank Prof. Christopher Yau, Dr. Eda Ozyigit, Albert Henry and Joe Farrington for their insightful guidance and discussion during this work.

\appendix

\section{Appendix}
\label{apd}

\begin{table*}[ht]
\centering
\caption{Model Hyperparameters. Ranges for tuning indicated by square braces. Final values from tuning under each disease acronym.}
\resizebox{0.95\linewidth}{!}{
    \begin{tabular}{lllllll}
    \toprule
    BERT Optimizer & Tuning & MI & T2D & HF & DM & RA \\
    \midrule
    Learning Rate & [$1\times 10^{-5}$, $1\times 10^{-4}$, $1\times 10^{-3}$] & $1\times 10^{-4}$ & $1\times 10^{-4}$ & $1\times 10^{-4}$ & $1\times 10^{-4}$ & $1\times 10^{-4}$\\
    Warm-Up Proportion & 0.1 \\
    Weight Decay & 0.001 \\
    \midrule
    BERT Model Hyperparameters &  \\
    \midrule
    Batch Size & 256 \\
    Hidden Layer Size & [120, 240, 360] & 360 & 360 & 360 & 360 & 360 \\
    Number Of Hidden Layers & [6, 10, 12] & 2 & 6 & 10 & 6 & 6 \\
    Hidden Dropout Probability & 0.2 \\
    Number Of Multi-Head Attention Layers & [6, 10, 12] & 12 & 12 & 12 & 12 & 12 \\
    Intermediate Layer Size In Transformer & [128, 256, 512] & 512 & 512 & 256 & 256 & 256 \\
    Number Of Attention Heads & 12 \\
    Multi-Head Attention Dropout Rate & 0.22 \\
    Parameter Weight Initializer Range & 0.02 \\
    Non-Linear Activation (Encoder \&   Pooler) & GELU \\
    \bottomrule
    \end{tabular}
}
\label{tab:bert_hparams}
\end{table*}

\subsection{Additional Anchor Learning}
\label{apd:additionaL_results}

\begin{figure*}[ht]
\centering
  %
    \subfloat{\label{fig:hf_noise}%
      \includegraphics[width=0.3\linewidth]{images/case_noise_411.2.png}}%
    \subfloat{\label{fig:dm_noise}%
      \includegraphics[width=0.3\linewidth]{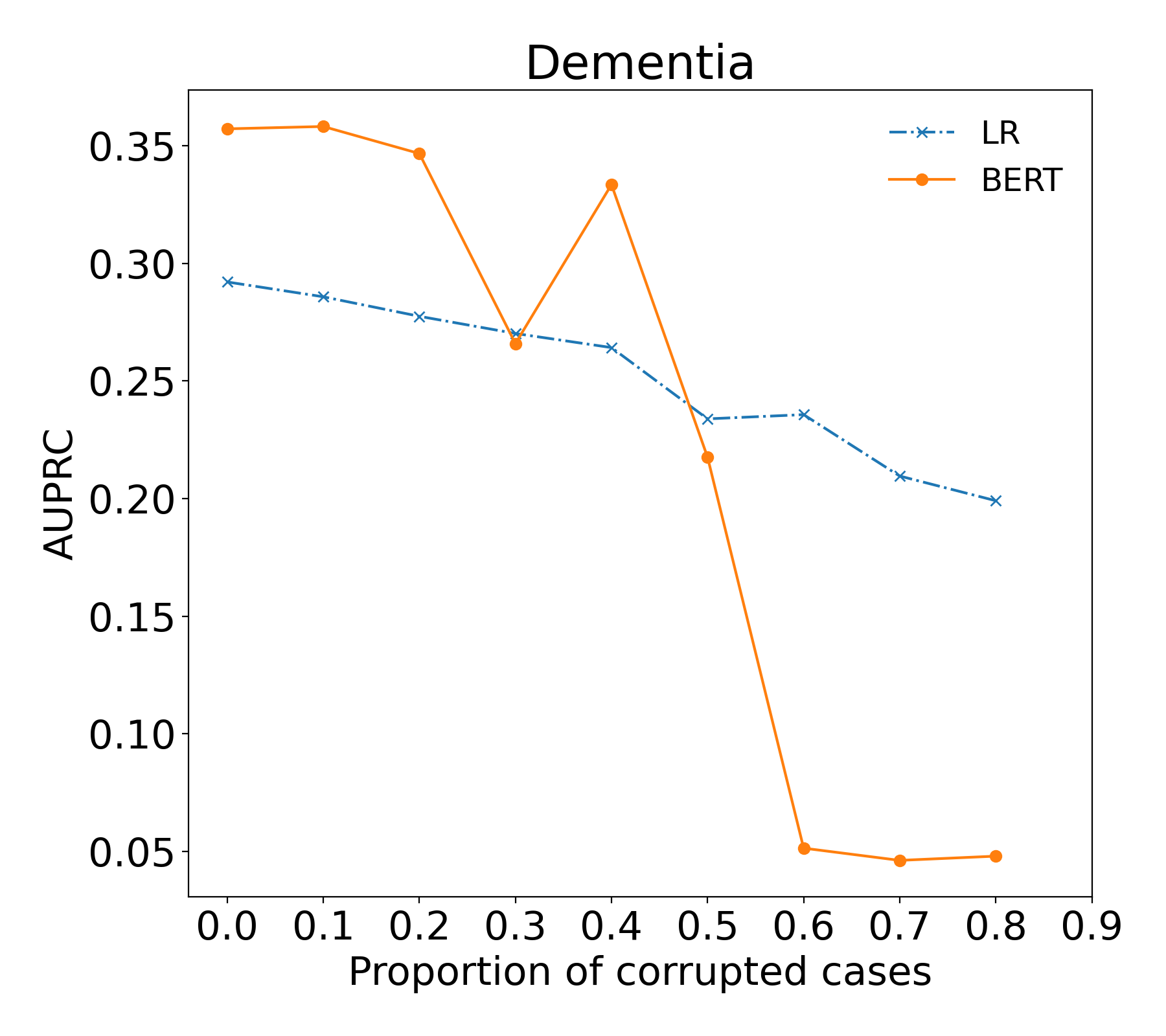}}%
    \subfloat{\label{fig:ra_noise}%
      \includegraphics[width=0.3\linewidth]{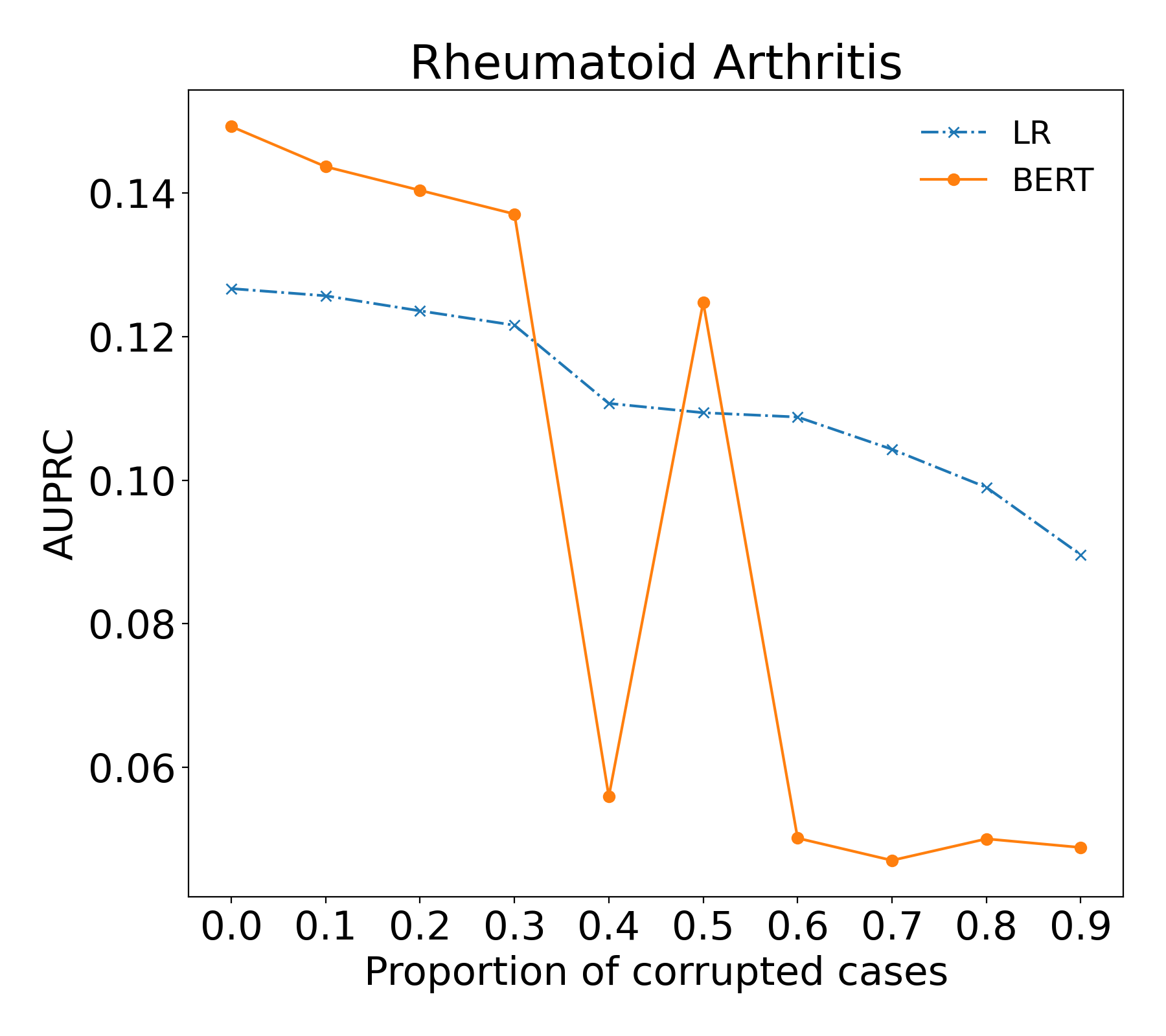}}%
  
  \label{fig:additional_noise}
  \caption{Area under precision recall curve for anchor variable classifiers as an increasing proportion of negative examples are flipped.}
\end{figure*}

\begin{figure*}[htbp]
\centering
    \subfloat[\label{fig:t2d_ablation_both}]{%
      \includegraphics[width=0.3\linewidth]{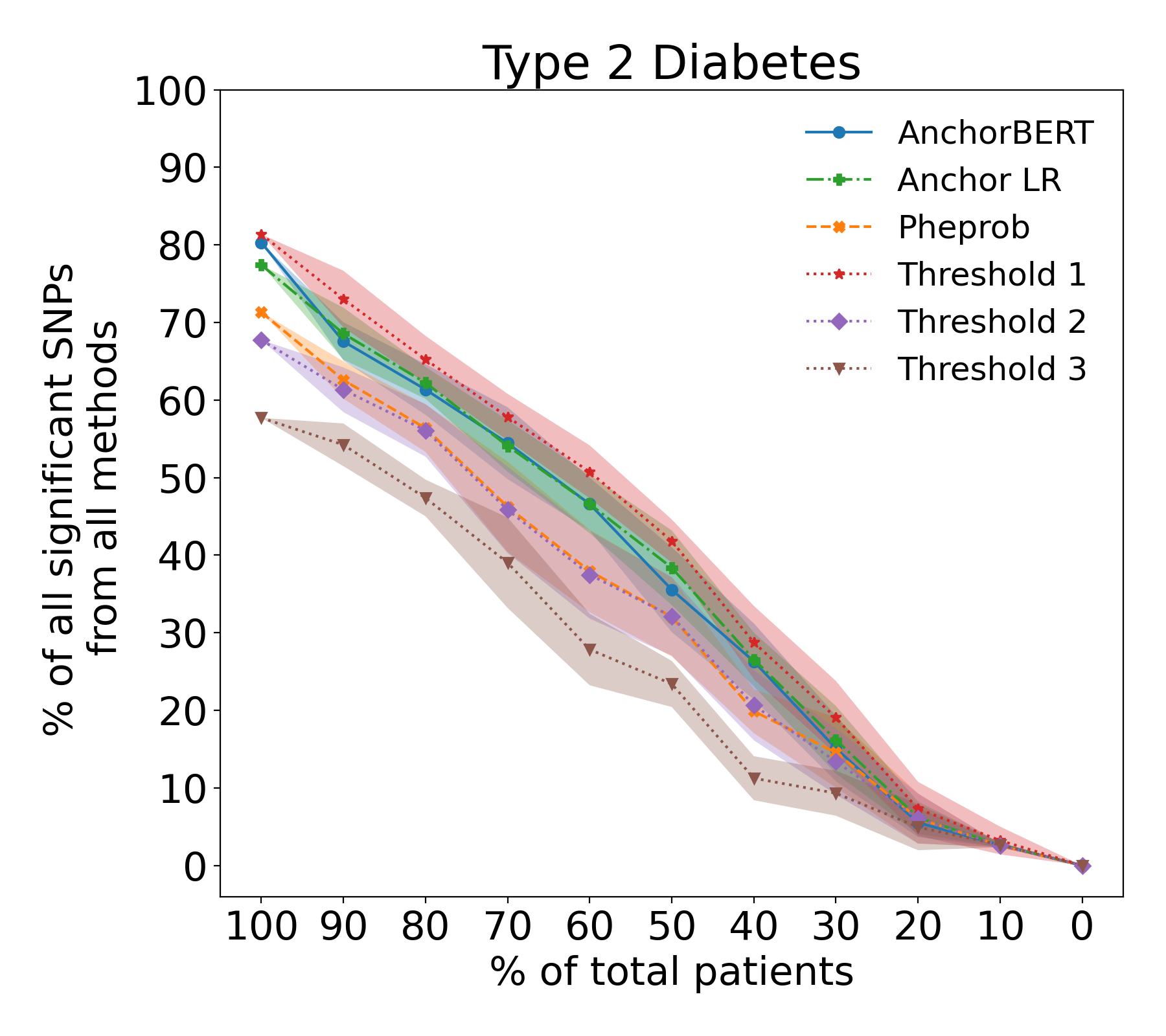}}%
    \subfloat[\label{fig:dm_ablation_both}]{%
      \includegraphics[width=0.3\linewidth]{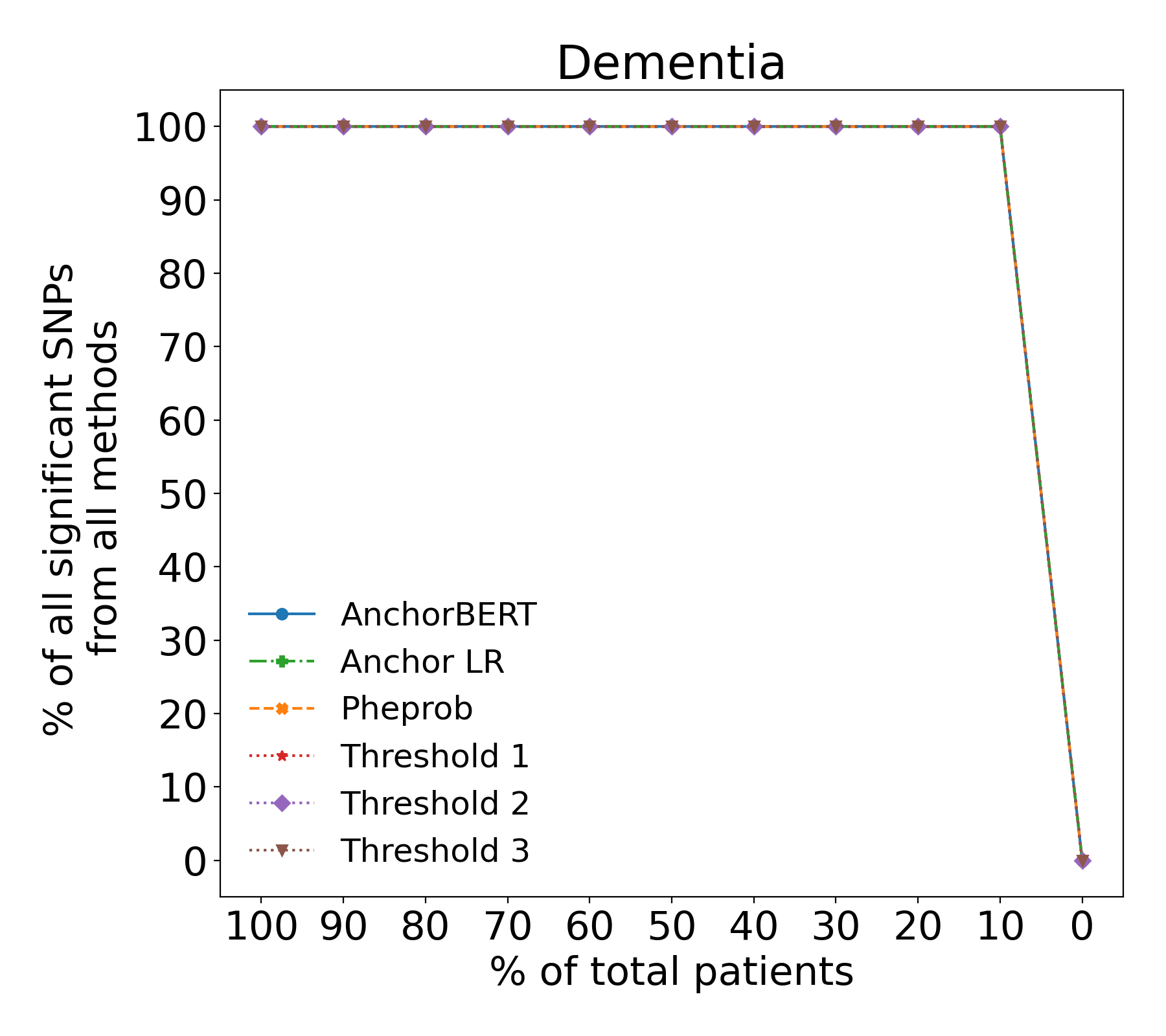}}%
    \subfloat[\label{fig:hf_ablation_both}]{%
      \includegraphics[width=0.3\linewidth]{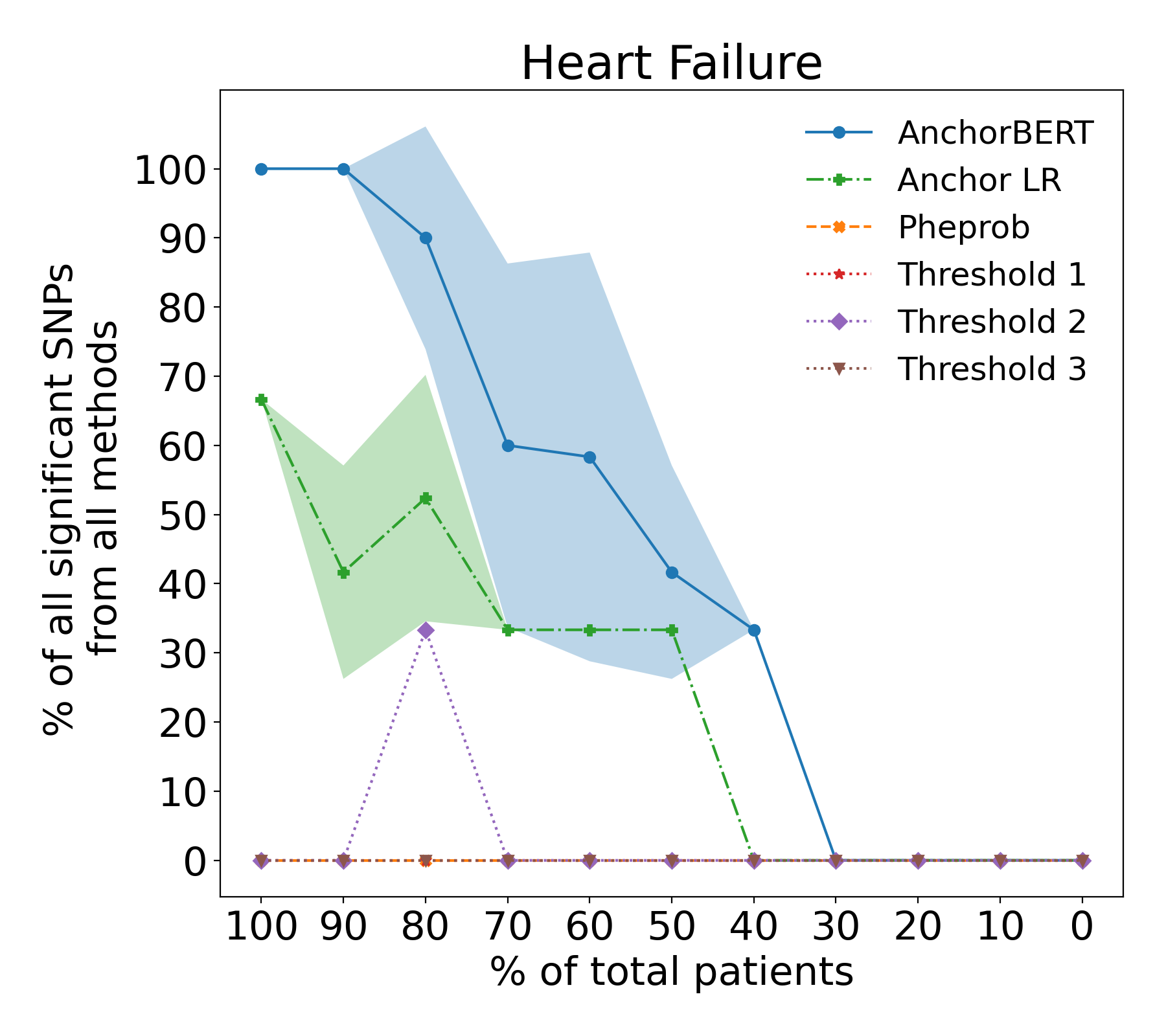}}%
    \qquad
    \subfloat[\label{fig:t2d_ablation_cases}]{%
      \includegraphics[width=0.3\linewidth]{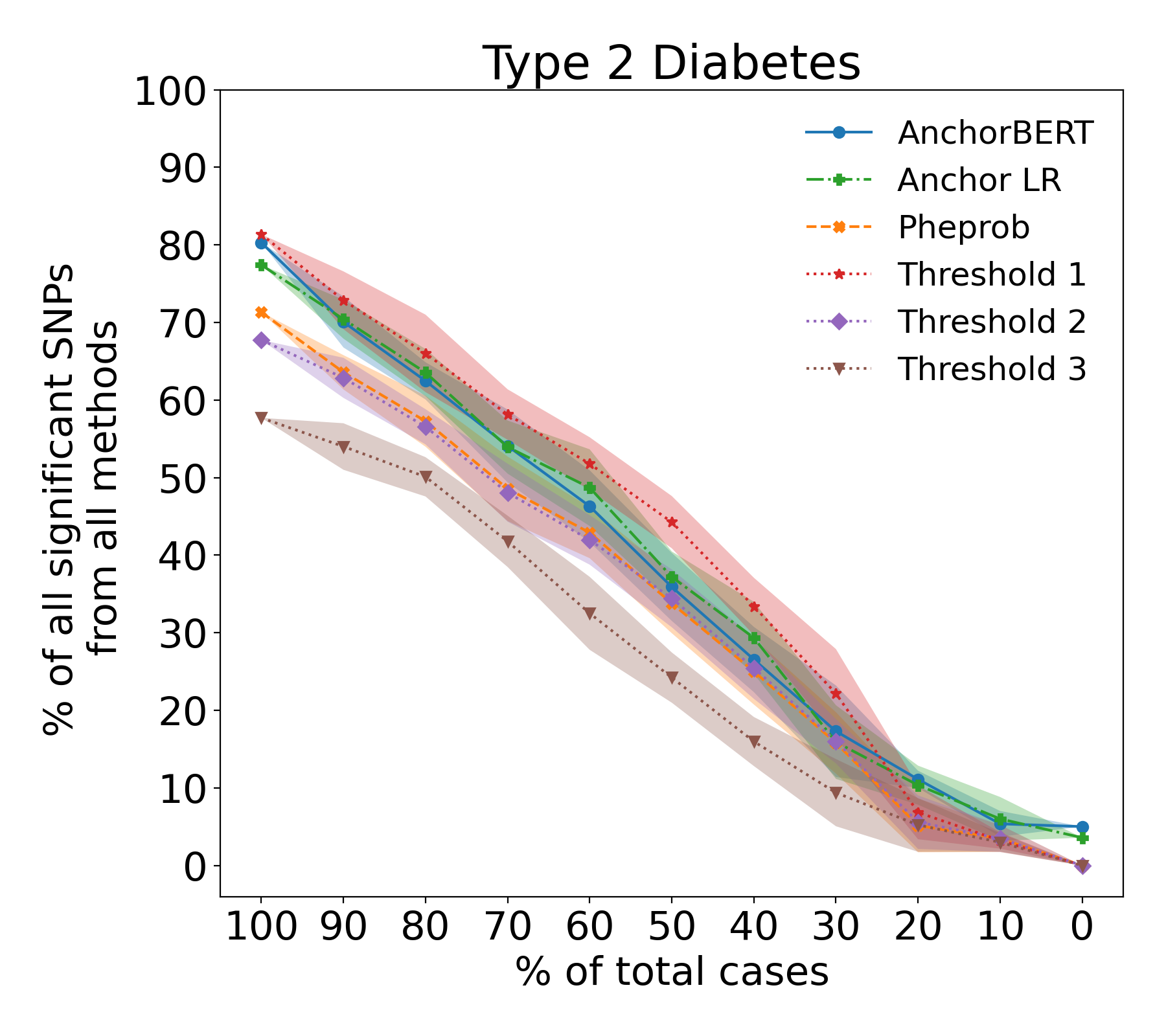}}
    \subfloat[\label{fig:dm_ablation_cases}]{%
      \includegraphics[width=0.3\linewidth]{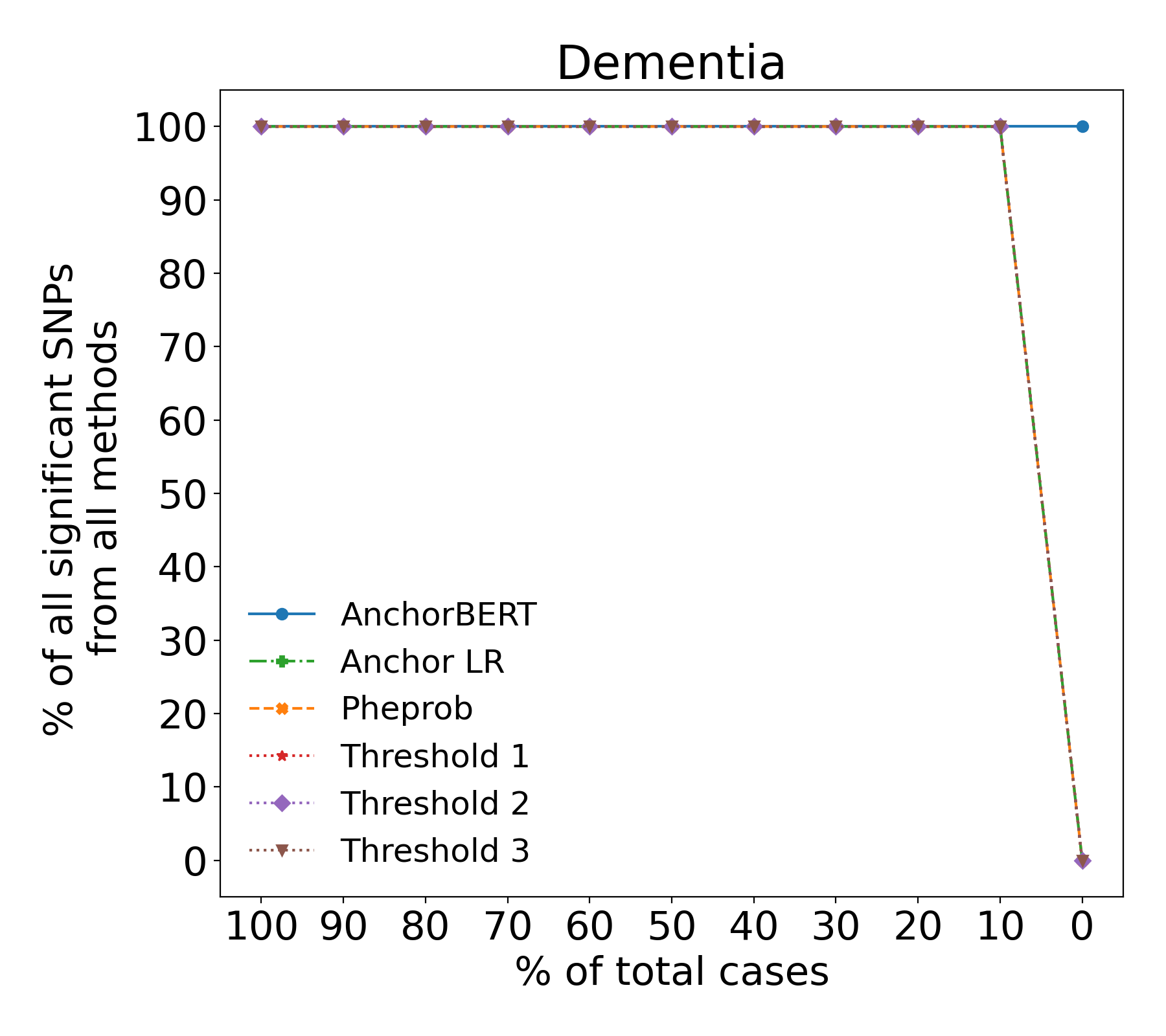}}
    \subfloat{\label{fig:hf_ablation_cases}%
      \includegraphics[width=0.3\linewidth]{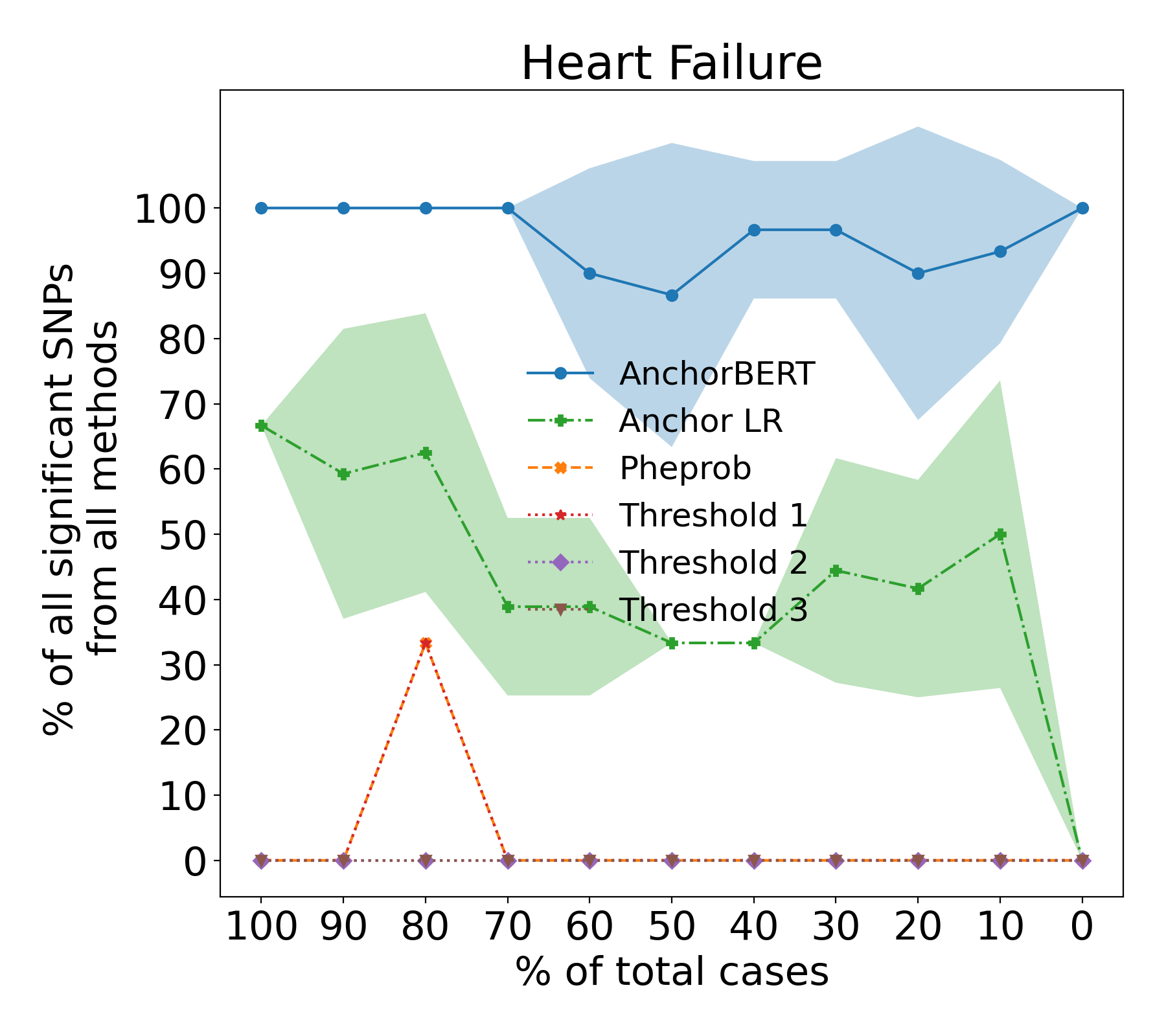}}
    \qquad
    \subfloat{\label{fig:ra_ablation_both}%
      \includegraphics[width=0.3\linewidth]{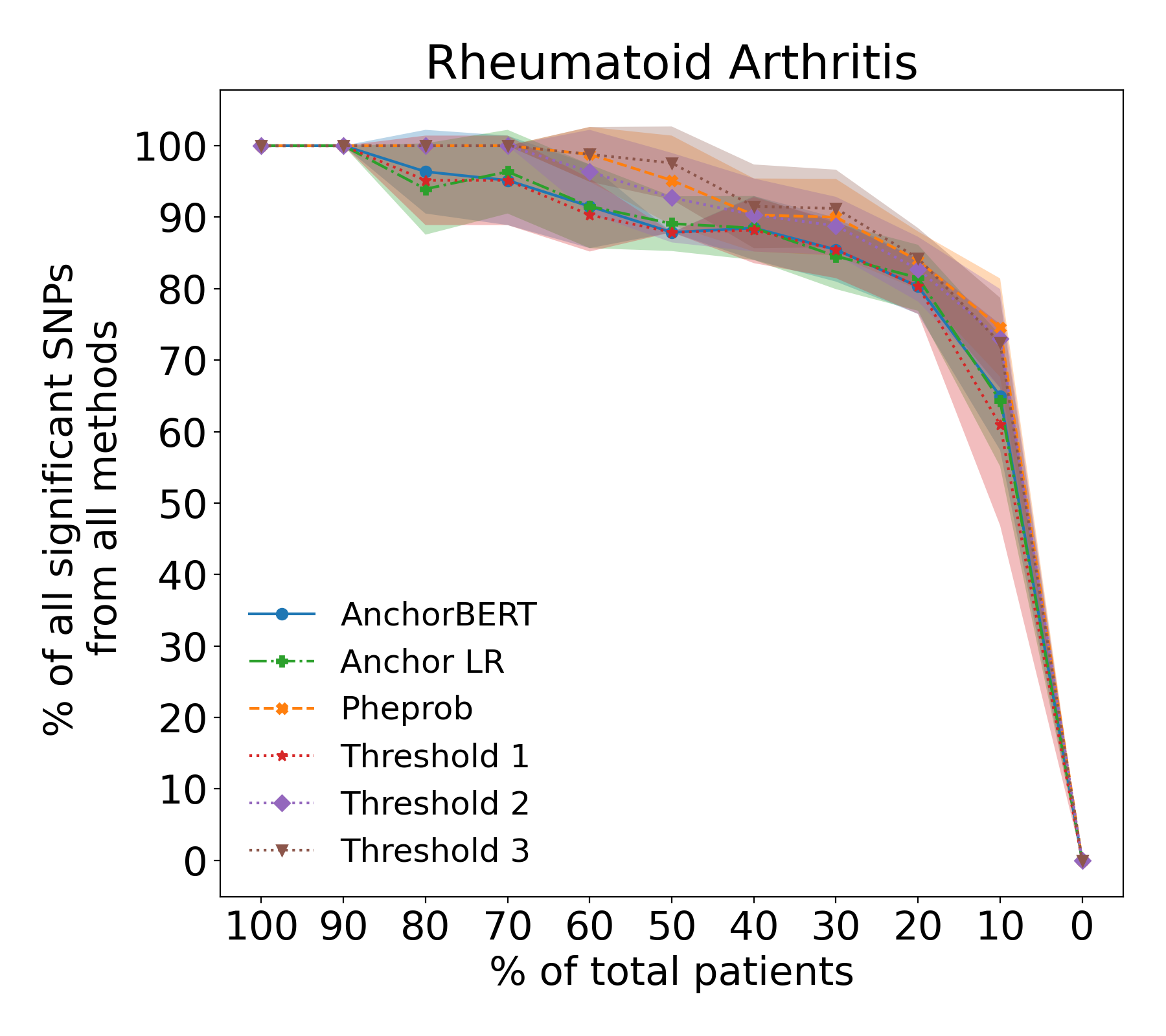}}%
    \subfloat{\label{fig:ra_ablation_cases}%
      \includegraphics[width=0.3\linewidth]{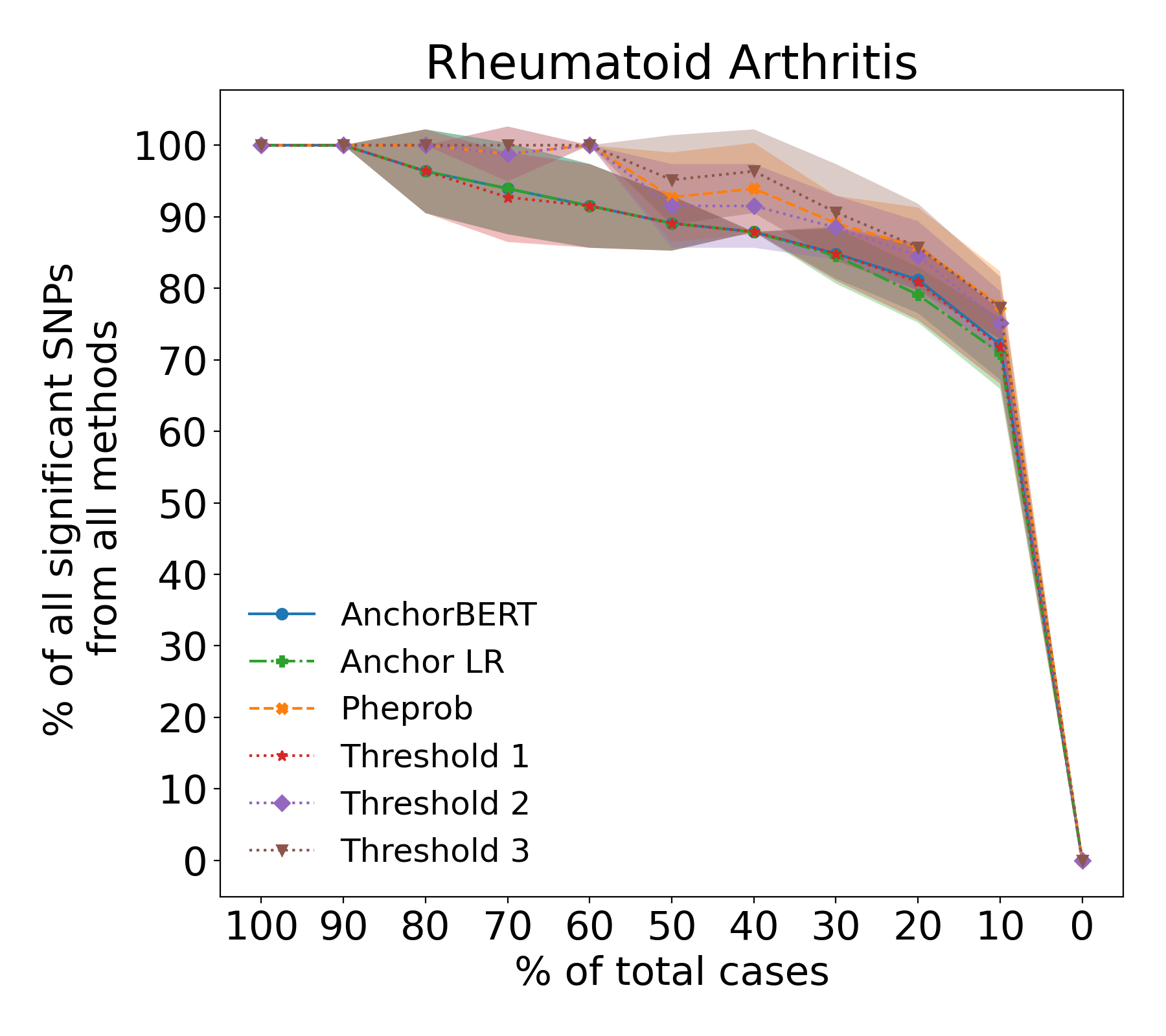}}
  \caption{Data ablation study for the remaining diseases for each phenotyping method. Shaded areas indicate one standard deviation of results across 10 trails.}
  \label{fig:additional_ablation}
\end{figure*}

\subsection{Additional GWAS Results}\label{apd:additional_gwas}

\begin{table*}[htbp]
\caption{Total SNPs found in our own GWAS full data experiments, found in the catalog and found after LD matching. There are often a greater number of GWAS matches than catalog matches as multiple SNPs in our analysis maybe in LD with the reported lead SNP in the catalog. The difference between matched and unmatched SNPs in our analysis may represent false positive association or as yet undiscovered genetic associations, further analysis is needed to determine the number of likely causal SNPs in this set.}
\centering
\begin{tabular}{llllll}
\toprule
 &  & \multicolumn{4}{c}{Total Significant SNPS} \\
Disease & Method & GWAS & of which found matches & Catalog & of which found matches \\
\midrule
\multirow{6}{*}{MI} & AnchorBERT & 1,396 & 903 & 128 & 44 \\
 & Anchor LR & 987 & 728 & 128 & 39 \\
 & Pheprob & 517 & 277 & 128 & 28 \\
 & Threshold-1 & 512 & 299 & 128 & 29 \\
 & Threshold-2 & 510 & 270 & 128 & 25 \\
 & Threshold-3 & 413 & 210 & 128 & 18 \\
 \cmidrule(r){2-6}
\multirow{6}{*}{T2D} & AnchorBERT & 8,763 & 8,631 & 1,758 & 266 \\
 & Anchor LR & 8,785 & 8,671 & 1,758 & 254 \\
 & Pheprob & 2,527 & 2,259 & 1,758 & 247 \\
 & Threshold-1 & 6,093 & 6,001 & 1,758 & 280 \\
 & Threshold-2 & 2,539 & 2,452 & 1,758 & 237 \\
 & Threshold-3 & 1,259 & 1,203 & 1,758 & 200 \\
 \cmidrule(r){2-6}
\multirow{6}{*}{RA} & AnchorBERT & 10,382 & 9,967 & 393 & 32 \\
 & Anchor LR & 9,867 & 9,448 & 393 & 32 \\
 & Pheprob & 11,943 & 11,411 & 393 & 32 \\
 & Threshold-1 & 9,884 & 9,519 & 393 & 32 \\
 & Threshold-2 & 11,877 & 11,350 & 393 & 32 \\
 & Threshold-3 & 10,074 & 9,593 & 393 & 32 \\
 \cmidrule(r){2-6}
\multirow{6}{*}{HF} & AnchorBERT & 155 & 152 & 42 & 3 \\
 & Anchor LR & 30 & 30 & 42 & 2 \\
 & Pheprob & 1 & 0 & 42 & 0 \\
 & Threshold-1 & 0 & 0 & 42 & 0 \\
 & Threshold-2 & 3 & 0 & 42 & 0 \\
 & Threshold-3 & 3 & 0 & 42 & 0 \\
 \cmidrule(r){2-6}
\multirow{6}{*}{DM} & AnchorBERT & 358 & 69 & 1 & 1 \\
 & Anchor LR & 338 & 69 & 1 & 1 \\
 & Pheprob & 298 & 68 & 1 & 1 \\
 & Threshold-1 & 325 & 70 & 1 & 1 \\
 & Threshold-2 & 301 & 68 & 1 & 1 \\
 & Threshold-3 & 286 & 68 & 1 & 1 \\
 \bottomrule
\end{tabular}
\label{tab:full_catalog_results}
\end{table*}

\bibliographystyle{plainnat}  
\bibliography{references_manual}

\end{document}